\documentclass[aps,prb,preprint,superscriptaddress,showpacs]{revtex4-1}

\usepackage{graphicx}% Include figure files

\begin{document}

%\preprint{}

\title{Transient photoluminescence enhancement as a probe of the
  structure of impurity-trapped excitons in CaF$_2$:Yb$^{2+}$}


\author{Michael F. Reid}
% \altaffiliation[Also: ]{MacDiarmid Institute for
%  Advanced Materials and Nanotechnology, University of Canterbury,
%  Christchurch, New Zealand}
 \affiliation{Department of Physics and Astronomy and MacDiarmid
  Institute for Advanced Materials and Nanotechnology, University of
  Canterbury, PB 4800, Christchurch 8140, New Zealand}

\author{Pubudu S. Senanayake}
\author{Jon-Paul R. Wells}
\email[Corresponding author: ]{jon-paul.wells@canterbury.ac.nz}
\affiliation{Department of Physics and Astronomy, University of
  Canterbury, PB 4800, Christchurch 8140, New Zealand}

\author{Giel Berden} \affiliation{FELIX Free Electron Laser Facility,
  FOM Institute for Plasmaphysics Rijnhuizen, PO Box 1207, 3430 BE,
  Nieuwegein, The Netherlands.}

% \author{Gabriele Bellocchi} 
\author{Andries Meijerink}
\affiliation{Debye Institute for NanoMaterials Science, University of
  Utrecht, P.O. Box 80000, TA 3508 Utrecht, The Netherlands}

\author{Alexander J. Salkeld} \affiliation{Department of Physics and Astronomy,
  University of Canterbury, PB 4800, Christchurch 8140, New Zealand}
\author{Chang-Kui Duan} \affiliation{Department of Physics, University
  of Science and Technology of China, Hefei 230026, China}

\author{Roger J. Reeves} 
%\altaffiliation[Also: ]{MacDiarmid Institute for
%  Advanced Materials and Nanotechnology, University of Canterbury,
%  Christchurch, New Zealand}
\affiliation{Department of Physics and Astronomy and MacDiarmid
  Institute for Advanced Materials and Nanotechnology, University of
  Canterbury, PB 4800, Christchurch 8140, New Zealand}
%\affiliation{Department of Physics and Astronomy, University of
%  Canterbury, PB 4800, Christchurch 8140, New Zealand}


\date{\today}

\begin{abstract}
  We demonstrate a direct measurement of the energy levels of
  impurity-trapped excitons in CaF$_2$:Yb$^{2+}$.  The radically
  different radiative decay rates of the lowest exciton state and higher
  excited states enable the generation of a transient photoluminescence
  enhancement measured via a two-step excitation process.  We observe
  sharp transitions arising from changes of state of localized
  electrons, broad bands associated with changes of state of delocalized
  electrons, and broad bands arising from trap liberation.

\end{abstract}


% 76.30.Kg 	Rare-earth ions and impurities 
% 71.35.-y 	Excitons and related phenomena
% 78.47.D- 	Time resolved spectroscopy (>1 psec) 
% 71.70.Ch 	Crystal and ligand fields 

%\keywords{}

\maketitle


%\subsection{}
%\subsubsection{}

% If in two-column mode, this environment will change to single-column
% format so that long equations can be displayed. Use
%\begin{widetext}
%\end{widetext}

The unique optical properties of rare-earth doped materials are
responsible for their crucial role in a wide variety of applications
such as fluorescent tubes, white light LEDs, lasers, fiber amplifiers,
and medical imaging.\cite{LiJa05,Ro07} The sharp-line optical
transitions within the $4f^N$ ($N=0$--$14$) ground configurations of
rare-earth ions have been extensively studied and may be accurately
modeled.\cite{LiJa05,Ro07,Ca89} Transitions involving excited
configurations such as $4f^{N-1}5d$ are crucial to many applications but
information about these states is limited because $4f^N \leftrightarrow
4f^{N-1}5d$ spectra consist of broad vibronic bands, yielding much less
information than $4f^N \leftrightarrow 4f^N$
spectra.\cite{PiReWeSoMe02,PiReBuMe02,BuRe07,KaUrRe07,PaDuTa08} Modern ab-initio
methods \cite{SaSeBa10a,SaSeBa10b} are capable of accurately calculating
excited-state electronic structure and bond-length variations, giving
good agreement with available broad-band $4f^N \leftrightarrow
4f^{N-1}5d$ spectra.  The ab-initio calculations \cite{SaSeBa10b}
indicate that transitions \emph{between} the excited states would give a
combination of sharp-line and broad vibronic transitions. We have
recently proposed that two-frequency excitation experiments would
provide more detailed information about the excited states and better
test the calculations.\cite{ReHuFrDuXiYi10}

Excited configurations of rare-earth ions are not restricted to
$4f^{N-1}5d$. Configurations involving charge transfer between the
rare-earth ion and other ions in the material are also important.
Theoretical \cite{SaSeBa10a} and experimental \cite{GrMa08,MaGrCaBeBo09}
studies suggest that non-radiative relaxation of the $4f^{N-1}5d$
configuration is often mediated by trapped exciton states, where the
excited electron is no longer completely localized on the rare-earth
ion.

In some Eu$^{2+}$ and Yb$^{2+}$ materials emission from excitonic states
may be observed.\cite{McPe85,Do03a} Excitonic emission in CaF$_2$ and SrF$_2$
doped with Yb$^{2+}$ has been the subject of several
studies.\cite{McPe85,MoCoPe89,MoCoPe91,PeJoMc07} The ground-state electronic
configuration of CaF$_2$:Yb$^{2+}$ is $4f^{14}$.  UV excitation can
promote one of the $4f$ electrons to a $5d$ orbital, giving the excited 
configuration $4f^{13}5d$.  The $5d$ electron rapidly becomes
delocalized over the next-nearest-neighbor Ca$^{2+}$ ions.  The
Yb$^{2+}$ is then effectively ionized to Yb$^{3+}$, with electronic
configuration $4f^{13}$, i.e.\ one $4f$ \emph{hole}. This trivalent ion
attracts the F$^-$ nearest neighbors more strongly than a divalent ion,
leading to a large contraction of bond length.  Emission from the
exciton states to the $4f^{14}$ ground state involves the reverse change
in bond length, and therefore a broad, structureless, red-shifted
vibrational emission band.\cite{McPe85,Do03a}   Recent ab-initio calculations
have given valuable insight into the quantum physics of exciton
formation.\cite{SaSeBa10a}   However, the broad bands provide no
detailed information and experimental information on the energy-level
structure of impurity-trapped excitons is largely deduced from indirect
measurements such as temperature dependencies,\cite{MoCoPe89,MoCoPe91}
pressure dependencies,\cite{GrMa08}  and photoconductivity.\cite{PeJoMc07}

In this work we report on an investigation of the internal structure of
impurity-trapped excitons using two-frequency measurements of
single-crystal CaF$_2$ doped with Yb$^{2+}$.  By applying IR radiation
to the crystal after exciting it in the UV we induce transitions between
exciton states. Since some of the exciton excited states have much
higher radiative rates than the lowest exciton state we can detect the
excited state absorption by monitoring photoluminescence enhancement.

CaF$_{2}$:Yb$^{2+}$ crystals were grown using the vertical Bridgmann
technique.  The UV component of our two-frequency excitation was from a
Quantronix TOPAS traveling-wave optical parametric amplifier (OPA)
providing 3 ps pulses tunable in the 250-400 nm region of interest in
this work at a repetition rate of 1 kHz. Pulsed infrared excitation was
achieved using the Dutch free electron laser (FEL) FELIX in
Nieuwegein. The IR output of FELIX consists of a 4-6 $\mu$s
macropulse at a repetition rate of 10 Hz, containing micropulses at 25
MHz. FELIX is continuously tunable from 3 to 250 $\mu$m. The OPA was
synchronized to the FEL and the electronic timing between the two lasers
could be varied. The UV and IR beams were spatially (but not temporally)
overlapped on the sample, held at cryogenic temperatures within an
Oxford instruments \emph{microstat} helium flow cryostat.  Visible
fluorescence was detected using a TRIAX 320 spectrometer equipped
with a C31034 photomultiplier.

%\section{Experimental Results}

Our results for 365 nm pulsed UV excitation of CaF$_2$:Yb$^{2+}$ are
consistent with previous work,\cite{MoCoPe89,MoCoPe91}  which reported
strongly red-shifted fluorescence having a single-exponential decay with
a lifetime of 15 ms at 4.2~K that reduces at higher temperatures.

\begin{figure} 
% multiply by 0.53 for preprint, 0.9 for reprint. 
\includegraphics[width=0.53\columnwidth]{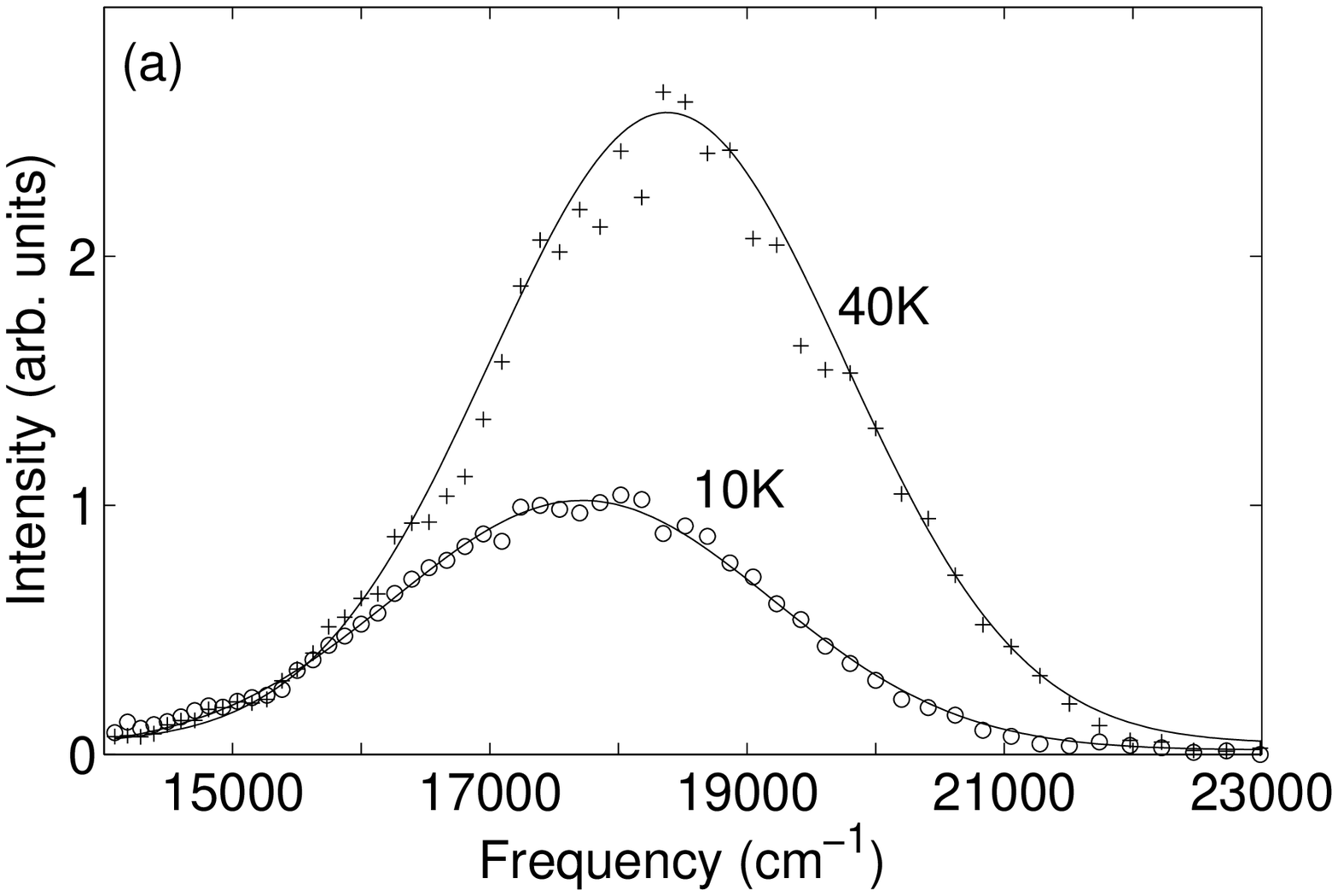}\\  
\includegraphics[width=0.53\columnwidth]{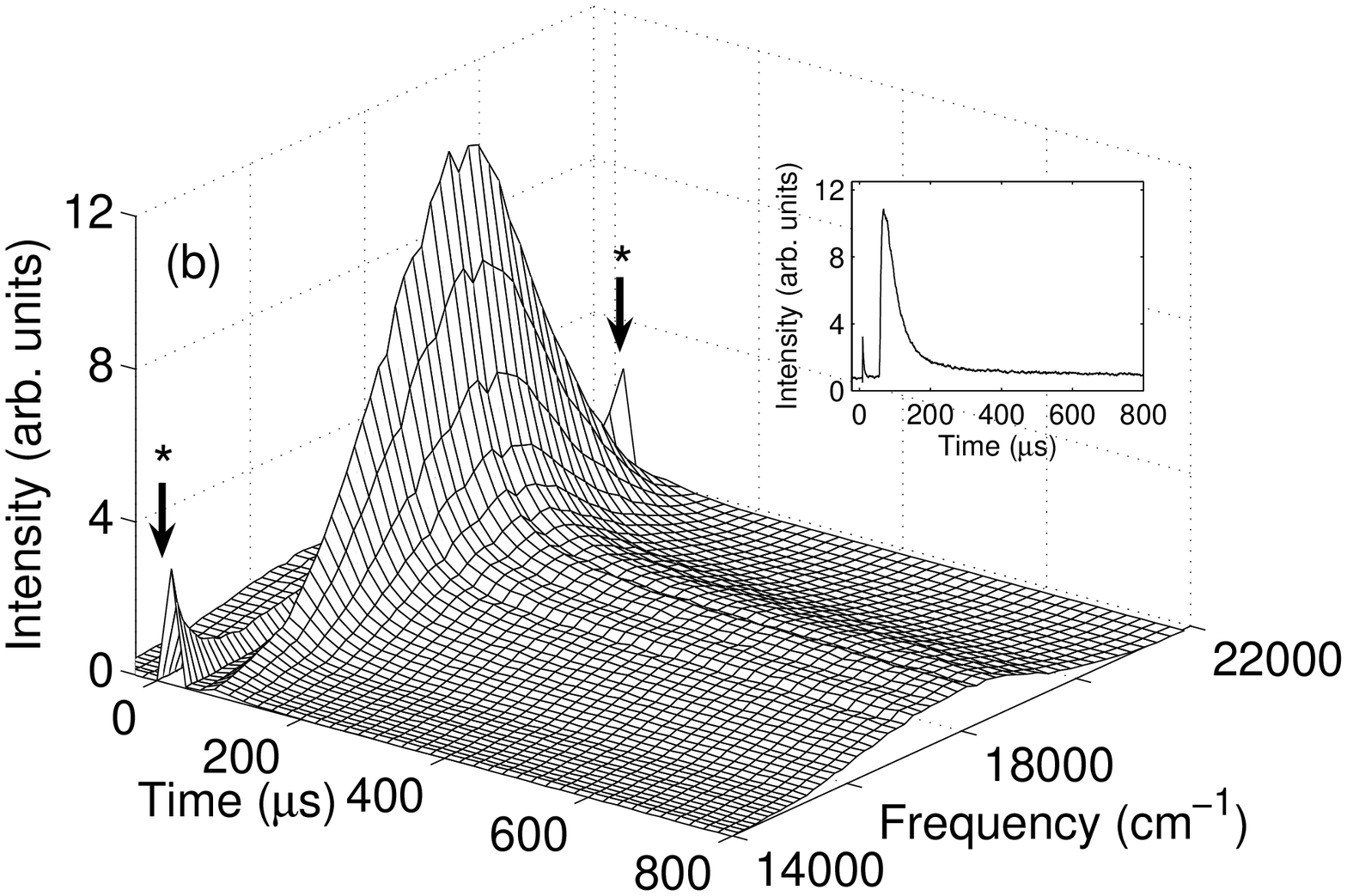}\\
\includegraphics[width=0.53\columnwidth]{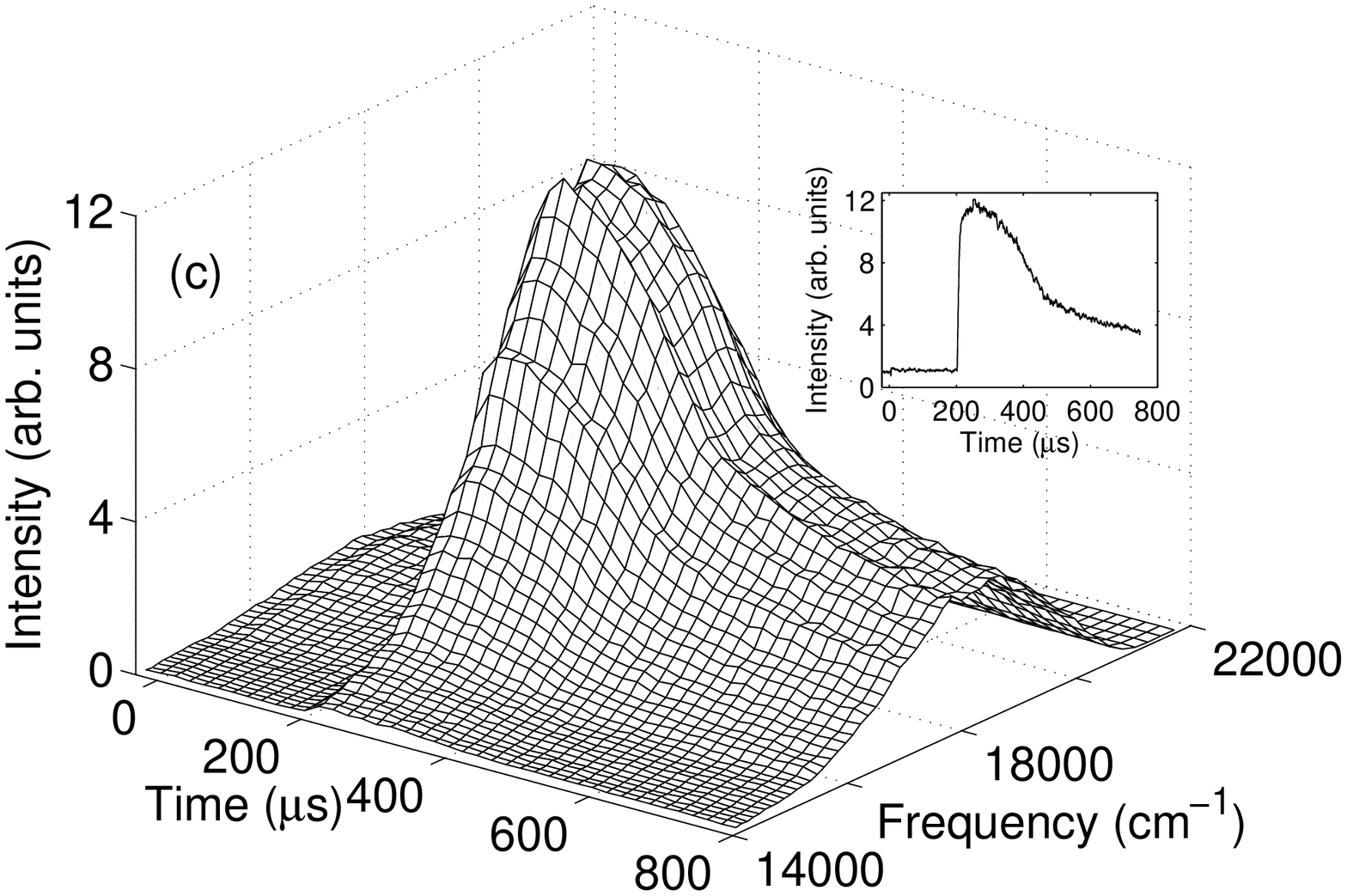}
\caption{\label{fig:FELemission} Fluorescence from CaF$_2$:Yb$^{2+}$ for
  excitation at 365 nm.  (a) Spectra at 10~K and 40~K 900 $\mu$s after
  UV excitation.  (b) Time evolution of 10~K emission spectrum excited
  with 12.1 $\mu$m (826 cm$^{-1}$) IR radiation following the UV
  pre-pulse. * Shows a combination of first- and second-order UV scatter
  and emission from minor Eu$^{2+}$ impurities.  (c) Time evolution of
  10~K emission spectrum excited with 16 $\mu$m (625 cm$^{-1}$) IR
  radiation following the UV prepulse. The (b) and (c) insets show the
  time profile at peak emission wavelength.}
\end{figure}

In Fig.~\ref{fig:FELemission}(a), the fluorescence spectra for
CaF$_{2}$:Yb$^{2+}$ at 10~K and 40~K are shown. At 10~K the band center
and width (FWHM) are 17720 and 3460 cm$^{-1}$ respectively. At 40~K the
intensity increases by a factor of 2.5 and the band center and width are
18380 and 3240 cm$^{-1}$ respectively. Previous analysis of the
temperature dependence \cite{MoCoPe89} suggests that the emission is
from two states whose  energy separation is about 40
cm$^{-1}$, with very different radiative lifetimes, 15~ms for the lower
state and 260~$\mu$s for the upper state. 
 
The shift in band center and change in bandwidth indicate that the two
excited states have different bond lengths. Emission bandwidths have
been previously used to calculate the change in bond length between the
lowest exciton state and the ground state as 0.17 \AA\ \cite{MoCoPe89},
much larger than bond-length changes for transitions between $4f^N$ and
$4f^{N-1}5d$ configurations.\cite{KaUrRe07,SaSeBa10b} The calculation is
approximate because it uses an effective phonon frequency. A frequency
of 325 cm$^{-1}$ gives a bond-length change of $-0.17$ \AA\ for the
lowest exciton state relative to the ground state (from our 10~K data)
and $-0.16$ \AA\ for the first excited exciton state (from our 40~K
data). The first excited state thus has a \emph{longer} bond length,
closer to the ground state $4f^{14}$ bond length.

We now present the results of our two-frequency transient measurements
at 10~K.  Figs.~\ref{fig:FELemission}(b) and \ref{fig:FELemission}(c)
show respectively the results of irradiating the system at a wavelength
of 12.1 $\mu$m (826 cm$^{-1}$) and 16 $\mu$m (625 cm$^{-1}$), delayed
from the UV excitation by 100--200 $\mu$s.  The application of the IR
pulse yields significant {\it enhancement} of the emission. Note that
the IR pulse on its own cannot induce this optical emission. The
enhancement occurs because we now populate excited excitonic states that
have significantly {\it higher} radiative rates.  An enhancement of the
total emission at long time-scales is also observed. We interpret this
as liberation of electrons from traps (discussed below), which is known
to occur under intense IR radiation.\cite{IzKlViBrGr07}

For 12.1 $\mu$m IR excitation there is a rise time of approximately 6
$\mu$s, and a decay time of 43 $\mu$s. The spectrum at the time
corresponding to the maximum emission intensity is similar to the 40~K
spectrum in Fig.~\ref{fig:FELemission}(a). The enhanced emission is
therefore likely to be a result of radiation from the state 40 cm$^{-1}$
above the lowest exciton state.  Further evidence for this conclusion is
that at 40~K the IR excitation gives only a small enhancement and no
change to the spectrum. The decay of the transient signal is much faster
than the estimated 260 $\mu$s radiative lifetime,\cite{MoCoPe89}  due to
non-radiative processes.\cite{MoCoPe91}

The spectral and temporal behavior for 16 $\mu$m excitation,
Fig.~\ref{fig:FELemission}(c), is very different. The time development
cannot be fitted by single exponential rise and decay times, but the
rise time is approximately 30 $\mu$s and the decay approximately 160
$\mu$s. At the time corresponding to the maximum emission intensity we
observe a mixture of low-energy and high-energy Gaussians.  Several
hundred $\mu$s after the IR excitation pulse the intensity of the
low-energy Gaussian is still significantly higher than before the IR
pulse, suggesting that there are significantly \emph{more} ions
radiating than before the IR excitation.

\begin{figure}
% multiply by 0.8 for preprint, 0.9 for reprint.
\includegraphics[width=0.8\columnwidth]{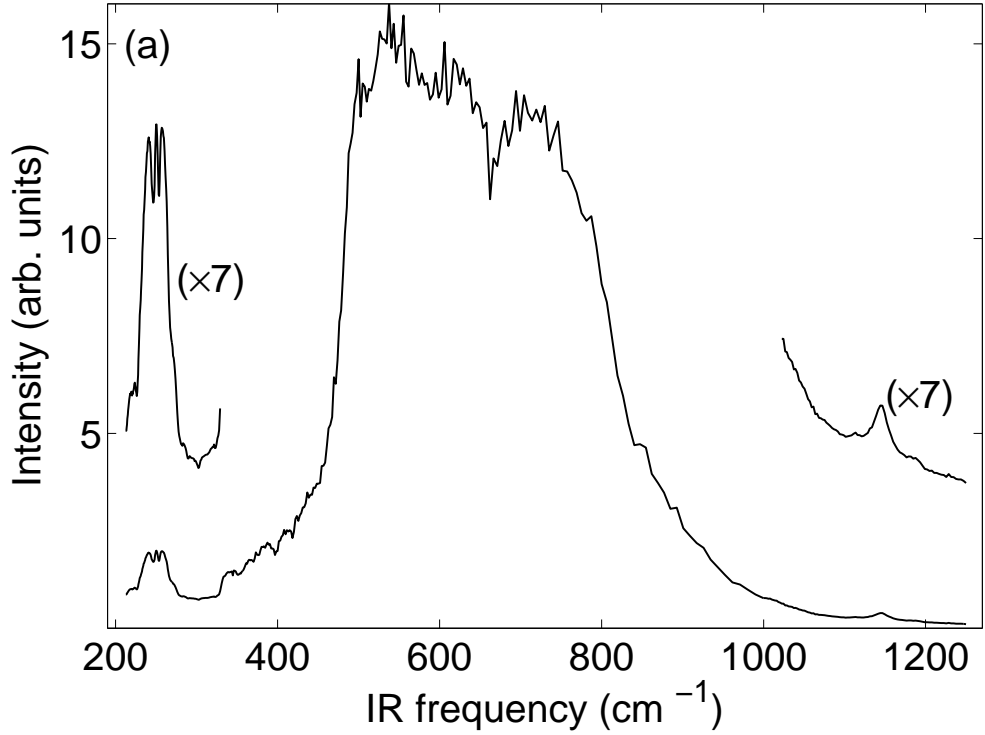}\\
\includegraphics[width=0.8\columnwidth]{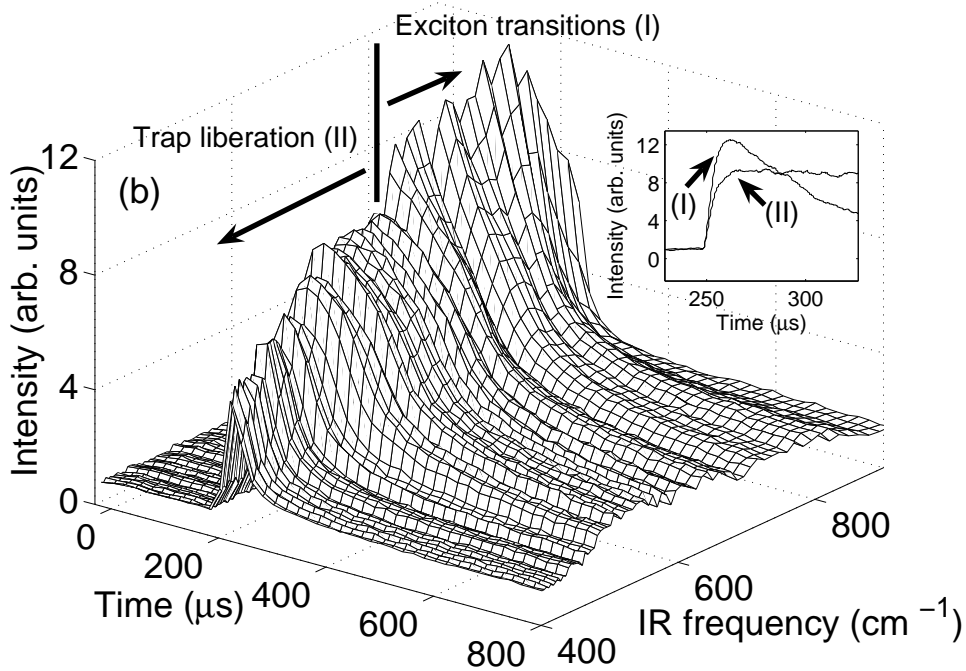}
\caption{\label{fig:FELspectrum} (a) 10~K IR excitation spectrum of
  CaF$_2$:Yb$^{2+}$ deduced by integrating emission enhancement over
  time.  The regions where sharp peaks are observed have been magnified
  and displaced for clarity. 
  (b) Time-evolution of the 10~K IR excitation spectrum.  Inset,
  rise time with (I) IR excitation at 825 cm$^{-1}$, (II) IR excitation
  at 625 cm$^{-1}$}
\end{figure}

The range of IR energies for which we have observed enhancement of the
visible emission is shown in the excitation spectrum of
Fig.~\ref{fig:FELspectrum}(a), where the photoluminescence transient is
time-integrated to give the signal.  The spectrum consists of a broad
band centered at 650 cm$^{-1}$ and two sharper peaks at 250 cm$^{-1}$
and 1145 cm$^{-1}$, with linewidths of 35 and 20 cm$^{-1}$
respectively. We note that dips in the broad band and the lower sharp
peak correlate with atmospheric absorption of the FEL radiation, which
can distort the lineshape, despite purging of the IR beam path with dry
N$_{2}$ gas.
 
The strong dependence of the temporal dynamics on the IR excitation
wavelength is further illustrated in figure \ref{fig:FELspectrum}(b).
The broad peak is actually two distinct regions, a higher-energy region
with fast rise and decay times, and a lower-energy region dominated by
slow rise and decay times. We attribute the long-lived signal in this
region to trap liberation processes. Whereas transitions within the
exciton transfer population from the lowest exciton state to a state
with a much higher radiative rate, trap liberation processes increase
the total number of excitons radiating. A phenomenological model of a
Coulomb trap having a threshold of 380 cm$^{-1}$ gives an asymmetric
spectrum with a width of approximately 400 cm$^{-1}$
(Ref.~\onlinecite{IzKlViBrGr07}, Eq.~(3)), which is consistent with the
width of the low-energy band.

The high-energy part of the band, from 650 to 950 cm$^{-1}$, and the
sharp lines, are assigned to transitions within the exciton.  Recall
that our exciton model is a Yb$^{3+}$ ion with one $4f$ hole ($4f^{13}$)
and a delocalized electron. It is the delocalized electron that can
affect bonding, so the vibronic broadening of the 650--950 cm$^{-1}$
band leads us to attribute this feature to transitions that change the
orbital of the delocalized electron and therefore the bond length.  The
width of this band is similar to the width of the phonon spectrum in
CaF$_2$,\cite{HaWiMaMc73} which implies that the bond-length change for
the transition is small, similar to the difference in bond-length
between the lowest two exciton states calculated above (0.01 \AA).  An
accurate calculation of the position and width of this band should be
possible with a detailed ab-initio approach, as in
Ref.~\onlinecite{SaSeBa10a}.

%%%%% Analysis of Sharp lines. 
 
The sharp excitation features observed at 250 and 1145 cm$^{-1}$ cannot
involve a change in bonding. We therefore assign them to changes in the
wavefunction of the localized $4f$ hole or the relative spin of the $4f$
hole and delocalized electron. The latter is associated with an exchange
Coulomb interaction. The calculations of Ref.~\onlinecite{SaSeBa10a} suggest
that the excitons involve a linear combination of $5d$ and $6s$ orbitals
with totally symmetric ($s$) character.  Unlike the SrCl$_2$:Yb$^{2+}$
system \cite{SaSeBa10b} detailed calculations for CaF$_2$:Yb$^{2+}$ are
not available.  However, we may model the sharp lines with a simple
semi-empirical model by constructing a ``crystal field'' Hamiltonian for
an $s$ electron and a $4f$ hole in a cubic crystal field,
\begin{eqnarray}\label{eq:Hcf}
H_\mathrm{cf} = \zeta A_\mathrm{so} 
  +  B^4 \left(C^4_0 + \sqrt{\frac{5}{14}} \left[C^4_4 + C^4_{-4}\right]
  \right) 
  \nonumber \\ 
  +  B^6 \left(C^6_0 - \sqrt{\frac{7}{2}}\left[C^6_4 + C^6_{-4}\right] \right)  
+ G^3({fs})   g_3({fs}) ,
\end{eqnarray}
where $\zeta$ is the spin-orbit interaction for the $4f$ electrons,
$B^4$ and $B^6$ are crystal-field parameters for the $4f$ electrons, and
$G^3({fs})$ the exchange interaction between the $4f$ and the
delocalized electron. Details of the Hamiltonian operators may be found
in Refs.~\onlinecite{Ca89, Cow81, PiReWeSoMe02, BuRe07, PaDuTa08}

\begin{figure}
% multiply by 0.8 for preprint, 0.9 for reprint.
\includegraphics[width=0.8\columnwidth]{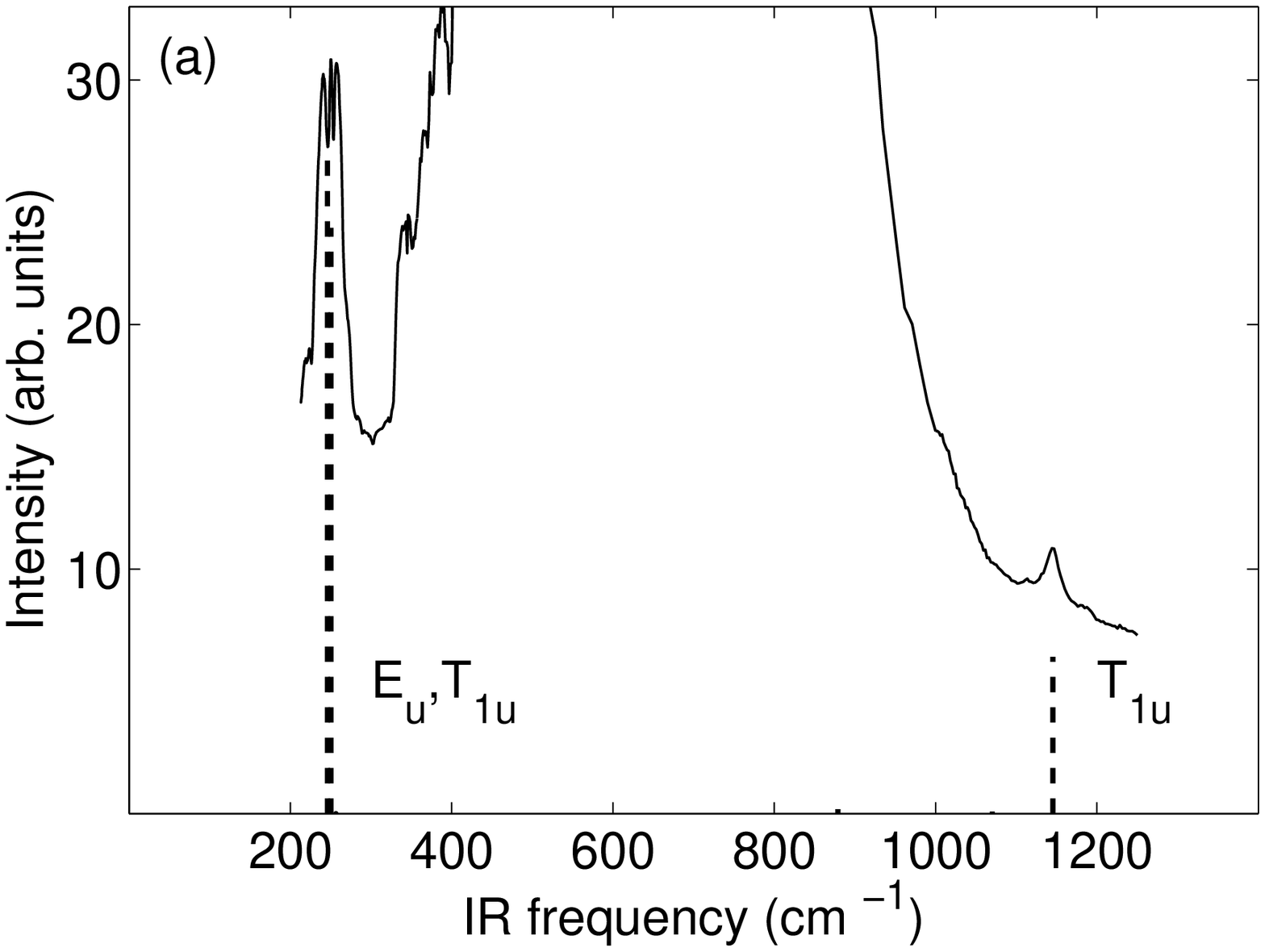}\\
\includegraphics[width=0.8\columnwidth]{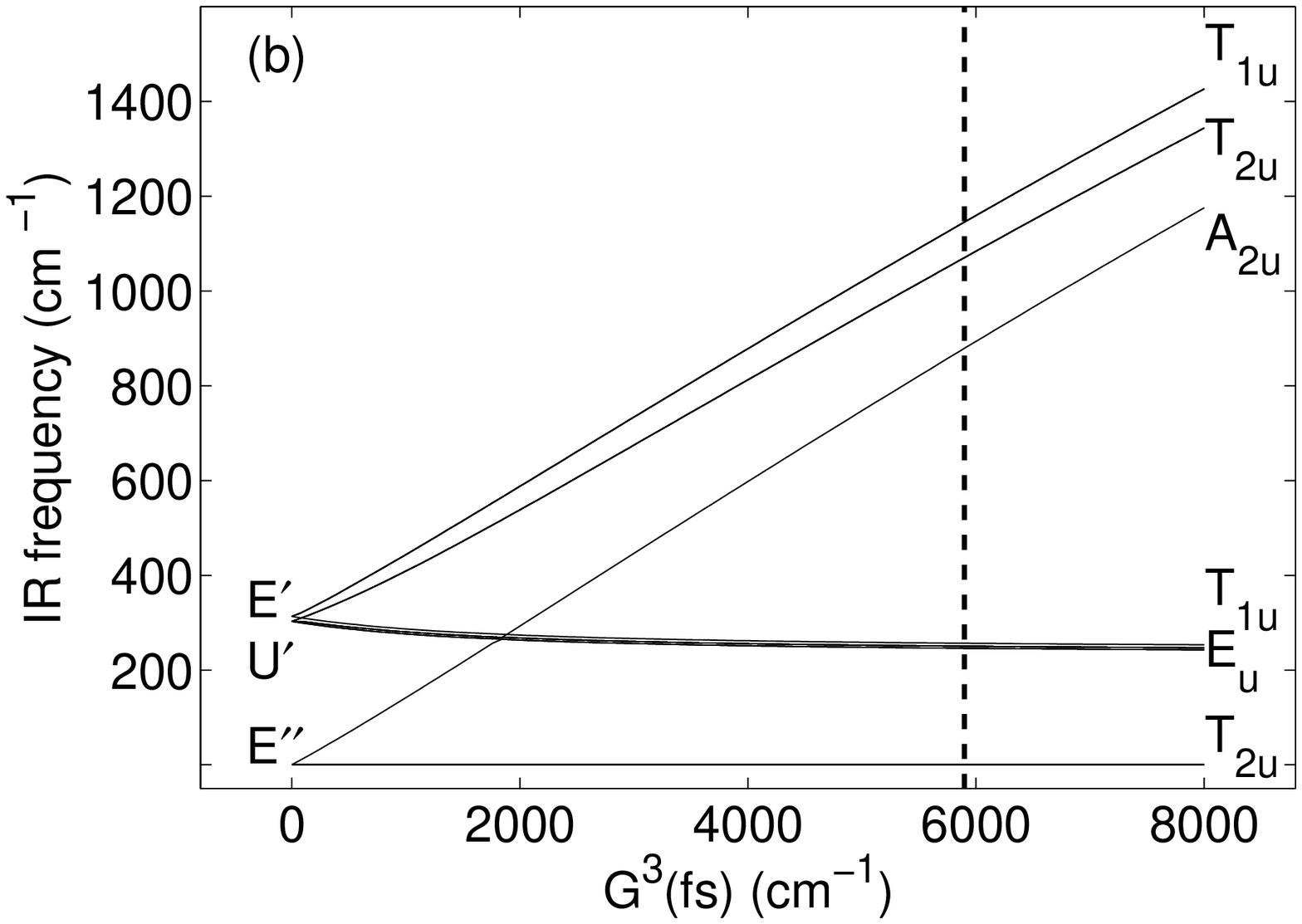}
\caption{\label{fig:ExchangeSplitting} (a) Calculated oscillator
  strengths (vertical lines) and experimental IR excitation spectrum. 
  (b) Dependence of the exciton energy levels on the exchange parameter
  $G^3({fs})$. The vertical dotted line is at the value of  $G^3({fs})$
  used in the calculation of Fig.~\ref{fig:ExchangeSplitting}(a).
%\\
%\emph{NUMBERS ON THE G3 GRAPH NEED TO DOUBLE, WITH THE EXPERIMENTAL
%  VALUE AT 5898.}
}
\end{figure}

Since we only observe two sharp lines it is not possible to fit all of
the parameters in Eq.~(\ref{eq:Hcf}). The spin-orbit parameter,
$\zeta$, was fixed at 2928 cm$^{-1}$, the value determined for Yb$^{3+}$
in LaF$_3$.\cite{Ca89} The ratio $B^6/B^4$ was fixed at $-0.35$, the
value obtained by Le{\'s}niak \cite{Les90} for Er$^{3+}$ in CaF$_2$.
With these assumptions the parameter values $B^4 = -800$ cm$^{-1}$ and
$G^3({fs}) = 5900$ cm$^{-1}$ give energy levels that match the observed
sharp lines.

Fig.~\ref{fig:ExchangeSplitting} illustrates our calculation. When the
$G^3({fs})$ parameter is set to zero the $^2F_{7/2}$ multiplet of
$4f^{13}$ splits into $E''$, $U'$ and $E'$ irreducible representations
of the octahedral group,\cite{HI89} as illustrated on the left side of
Fig.~\ref{fig:ExchangeSplitting}(b).  As $G^3({fs})$ is increased
these states couple with $^2S$ (from the $s$ electron).  In cubic symmetry
only magnetic-dipole transitions are allowed.  Calculated absorption
oscillator strengths are compared with experiment in
\ref{fig:ExchangeSplitting}(a). Note that several transitions are
forbidden or calculated to be very small.

The $B^4$ parameter is similar to the value of $-725$ cm$^{-1}$ for the
$4f^{13}5d$ configuration of Yb$^{2+}$ in SrCl$_2$.\cite{PaDuTa08}  A
Hartree-Fock calculation using the atomic code of Cowan \cite{Cow81}
gives $G^3({fs}) = 3167$ cm$^{-1}$ for the $4f^{13}6s$ configuration of
Yb$^{2+}$. This is smaller than our experimental value.  However, the
calculations of Ref.~\onlinecite{SaSeBa10a} suggest that the exciton will
contain mixtures of $5d$ and $6s$ orbitals. The exchange splitting for
$4f^{13}5d$ is over 2000 cm$^{-1}$,\cite{PaDuTa08} so our observed
splitting of about 1145 cm$^{-1}$ for the exciton, where the electron is
in a more delocalized orbital than $5d$, is reasonable.

% \section{Conclusions}

In conclusion, we have demonstrated that it is possible to probe the
internal structure and dynamics of exciton states, and trap-liberation
processes, in rare-earth materials with a two-frequency selective
fluorescence enhancement technique. A novel feature of our work is the
observation of sharp-line transitions within the exciton spectrum. A
quantitative account of these sharp line features has been obtained
using a parametrized crystal-field model.  Our observations provide a
basis and motivation for detailed ab-initio calculations as in
Ref.~\onlinecite{SaSeBa10a} and for investigation of other rare-earth
excitonic systems using the same experimental approach.

%\section{Acknowledgements}

\bigskip

This work was supported by the Marsden fund of the Royal Society of New
Zealand, Grant No.\ 09-UOC-080. We thank the Dutch FOM organization for
providing FELIX beamtime and the FELIX staff for their assistance.  Mr
P. S. Senanayake acknowledges support from the University of
Canterbury. C.-K. Duan acknowledges support from the Natural Science
Foundation of China, Grant No.\ 11074315.

% figures should be put into the text as floats.
% Use the graphics or graphicx packages (distributed with LaTeX2e)
% and the \includegraphics macro defined in those packages.
% See the LaTeX Graphics Companion by Michel Goosens, Sebastian Rahtz,
{% entire page. There is no need to do explicit centering.


\begin{thebibliography}{23}%
\makeatletter
\providecommand \@ifxundefined [1]{%
 \@ifx{#1\undefined}
}%
\providecommand \@ifnum [1]{%
 \ifnum #1\expandafter \@firstoftwo
 \else \expandafter \@secondoftwo
 \fi
}%
\providecommand \@ifx [1]{%
 \ifx #1\expandafter \@firstoftwo
 \else \expandafter \@secondoftwo
 \fi
}%
\providecommand \natexlab [1]{#1}%
\providecommand \enquote  [1]{``#1''}%
\providecommand \bibnamefont  [1]{#1}%
\providecommand \bibfnamefont [1]{#1}%
\providecommand \citenamefont [1]{#1}%
\providecommand \href@noop [0]{\@secondoftwo}%
\providecommand \href [0]{\begingroup \@sanitize@url \@href}%
\providecommand \@href[1]{\@@startlink{#1}\@@href}%
\providecommand \@@href[1]{\endgroup#1\@@endlink}%
\providecommand \@sanitize@url [0]{\catcode `\\12\catcode `\$12\catcode
  `\&12\catcode `\#12\catcode `\^12\catcode `\_12\catcode `\%12\relax}%
\providecommand \@@startlink[1]{}%
\providecommand \@@endlink[0]{}%
\providecommand \url  [0]{\begingroup\@sanitize@url \@url }%
\providecommand \@url [1]{\endgroup\@href {#1}{\urlprefix }}%
\providecommand \urlprefix  [0]{URL }%
\providecommand \Eprint [0]{\href }%
\@ifxundefined \urlstyle {%
  \providecommand \doi  [0]{\begingroup \@sanitize@url \@doi}%
  \providecommand \@doi [1]{\endgroup \@@startlink {\doibase
  #1}doi:\discretionary {}{}{}#1\@@endlink }%
}{%
  \providecommand \doi  [0]{doi:\discretionary{}{}{}\begingroup
  \urlstyle{rm}\Url }%
}%
\providecommand \doibase [0]{http://dx.doi.org/}%
\providecommand \Doi [0]{\begingroup \@sanitize@url \@Doi }%
\providecommand \@Doi  [1]{\endgroup\@@startlink{\doibase#1}\@@Doi}%
\providecommand \@@Doi [1]{#1\@@endlink}%
\providecommand \selectlanguage [0]{\@gobble}%
\providecommand \bibinfo  [0]{\@secondoftwo}%
\providecommand \bibfield  [0]{\@secondoftwo}%
\providecommand \translation [1]{[#1]}%
\providecommand \BibitemOpen [0]{}%
\providecommand \bibitemStop [0]{}%
\providecommand \bibitemNoStop [0]{.\EOS\space}%
\providecommand \EOS [0]{\spacefactor3000\relax}%
\providecommand \BibitemShut  [1]{\csname bibitem#1\endcsname}%
%</preamble>
\bibitem [{\citenamefont {Liu}\ and\ \citenamefont {Jacquier}(2005)}]{LiJa05}%
  \BibitemOpen
  \bibinfo {editor} {\bibfnamefont {G.~K.}\ \bibnamefont {Liu}}\ and\ \bibinfo
  {editor} {\bibfnamefont {B.}~\bibnamefont {Jacquier}},\ eds.,\ \href@noop {}
  {\emph {\bibinfo {title} {Properties of Rare Earths in Optical Materials}}}\
  (\bibinfo  {publisher} {Springer, Berlin},\ \bibinfo {year}
  {2005})\BibitemShut {NoStop}%
\bibitem [{\citenamefont {Ronda}(2007)}]{Ro07}%
  \BibitemOpen
  \bibinfo {editor} {\bibfnamefont {C.}~\bibnamefont {Ronda}},\ ed.,\
  \href@noop {} {\emph {\bibinfo {title} {Luminescence: From Theory to
  Applications}}}\ (\bibinfo  {publisher} {Wiley-VCH},\ \bibinfo {address}
  {Weinheim},\ \bibinfo {year} {2007})\BibitemShut {NoStop}%
\bibitem [{\citenamefont {Carnall}\ \emph {et~al.}(1989)\citenamefont
  {Carnall}, \citenamefont {Goodman}, \citenamefont {Rajnak},\ and\
  \citenamefont {Rana}}]{Ca89}%
  \BibitemOpen
  \bibfield  {author} {\bibinfo {author} {\bibfnamefont {W.~T.}\ \bibnamefont
  {Carnall}}, \bibinfo {author} {\bibfnamefont {G.~L.}\ \bibnamefont
  {Goodman}}, \bibinfo {author} {\bibfnamefont {K.}~\bibnamefont {Rajnak}}, \
  and\ \bibinfo {author} {\bibfnamefont {R.~S.}\ \bibnamefont {Rana}},\
  }\href@noop {} {\bibfield  {journal} {\bibinfo  {journal} {J. Chem. Phys.},\
  }\textbf {\bibinfo {volume} {90}},\ \bibinfo {pages} {3443} (\bibinfo {year}
  {1989})}\BibitemShut {NoStop}%
\bibitem [{\citenamefont {van Pieterson}\ \emph
  {et~al.}(2002){\natexlab{a}}\citenamefont {van Pieterson}, \citenamefont
  {Reid}, \citenamefont {Wegh}, \citenamefont {Soverna},\ and\ \citenamefont
  {Meijerink}}]{PiReWeSoMe02}%
  \BibitemOpen
  \bibfield  {author} {\bibinfo {author} {\bibfnamefont {L.}~\bibnamefont {van
  Pieterson}}, \bibinfo {author} {\bibfnamefont {M.~F.}\ \bibnamefont {Reid}},
  \bibinfo {author} {\bibfnamefont {R.~T.}\ \bibnamefont {Wegh}}, \bibinfo
  {author} {\bibfnamefont {S.}~\bibnamefont {Soverna}}, \ and\ \bibinfo
  {author} {\bibfnamefont {A.}~\bibnamefont {Meijerink}},\ }\href@noop {}
  {\bibfield  {journal} {\bibinfo  {journal} {Phys. Rev. B},\ }\textbf
  {\bibinfo {volume} {65}},\ \bibinfo {pages} {045113} (\bibinfo {year}
  {2002}{\natexlab{a}})}\BibitemShut {NoStop}%
\bibitem [{\citenamefont {van Pieterson}\ \emph
  {et~al.}(2002){\natexlab{b}}\citenamefont {van Pieterson}, \citenamefont
  {Reid}, \citenamefont {Burdick},\ and\ \citenamefont
  {Meijerink}}]{PiReBuMe02}%
  \BibitemOpen
  \bibfield  {author} {\bibinfo {author} {\bibfnamefont {L.}~\bibnamefont {van
  Pieterson}}, \bibinfo {author} {\bibfnamefont {M.~F.}\ \bibnamefont {Reid}},
  \bibinfo {author} {\bibfnamefont {G.~W.}\ \bibnamefont {Burdick}}, \ and\
  \bibinfo {author} {\bibfnamefont {A.}~\bibnamefont {Meijerink}},\ }\href@noop
  {} {\bibfield  {journal} {\bibinfo  {journal} {Phys. Rev. B},\ }\textbf
  {\bibinfo {volume} {65}},\ \bibinfo {pages} {045114} (\bibinfo {year}
  {2002}{\natexlab{b}})}\BibitemShut {NoStop}%
\bibitem [{\citenamefont {Burdick}\ and\ \citenamefont {Reid}(2007)}]{BuRe07}%
  \BibitemOpen
  \bibfield  {author} {\bibinfo {author} {\bibfnamefont {G.~W.}\ \bibnamefont
  {Burdick}}\ and\ \bibinfo {author} {\bibfnamefont {M.~F.}\ \bibnamefont
  {Reid}},\ }in\ \href@noop {} {\emph {\bibinfo {booktitle} {Handbook on the
  Physics and Chemistry of the Rare Earths}}},\ Vol.~\bibinfo {volume} {37},\
  \bibinfo {editor} {edited by\ \bibinfo {editor} {\bibfnamefont {K.~A.}\
  \bibnamefont {{Gschneidner Jr.}}}, \bibinfo {editor} {\bibfnamefont {J.~C.}\
  \bibnamefont {Bunzli}}, \ and\ \bibinfo {editor} {\bibfnamefont {V.~K.}\
  \bibnamefont {Percharsky}}}\ (\bibinfo  {publisher} {North Holland},\
  \bibinfo {year} {2007})\ Chap.\ \bibinfo {chapter} {232}, pp.\ \bibinfo
  {pages} {61 -- 91}\BibitemShut {NoStop}%
\bibitem [{\citenamefont {Karbowiak}\ \emph {et~al.}(2007)\citenamefont
  {Karbowiak}, \citenamefont {Urbanowicz},\ and\ \citenamefont
  {Reid}}]{KaUrRe07}%
  \BibitemOpen
  \bibfield  {author} {\bibinfo {author} {\bibfnamefont {M.}~\bibnamefont
  {Karbowiak}}, \bibinfo {author} {\bibfnamefont {A.}~\bibnamefont
  {Urbanowicz}}, \ and\ \bibinfo {author} {\bibfnamefont {M.~F.}\ \bibnamefont
  {Reid}},\ }\href@noop {} {\bibfield  {journal} {\bibinfo  {journal} {Phys.
  Rev. B},\ }\textbf {\bibinfo {volume} {76}},\ \bibinfo {pages} {115125}
  (\bibinfo {year} {2007})}\BibitemShut {NoStop}%
\bibitem [{\citenamefont {Pan}\ \emph {et~al.}(2008)\citenamefont {Pan},
  \citenamefont {Duan},\ and\ \citenamefont {Tanner}}]{PaDuTa08}%
  \BibitemOpen
  \bibfield  {author} {\bibinfo {author} {\bibfnamefont {Z.}~\bibnamefont
  {Pan}}, \bibinfo {author} {\bibfnamefont {C.~K.}\ \bibnamefont {Duan}}, \
  and\ \bibinfo {author} {\bibfnamefont {P.~A.}\ \bibnamefont {Tanner}},\
  }\href@noop {} {\bibfield  {journal} {\bibinfo  {journal} {Phys. Rev. B},\
  }\textbf {\bibinfo {volume} {77}},\ \bibinfo {pages} {085114} (\bibinfo
  {year} {2008})}\BibitemShut {NoStop}%
\bibitem [{\citenamefont {S\'anchez-Sanz}\ \emph
  {et~al.}(2010){\natexlab{a}}\citenamefont {S\'anchez-Sanz}, \citenamefont
  {Seijo},\ and\ \citenamefont {Barandiar\'an}}]{SaSeBa10a}%
  \BibitemOpen
  \bibfield  {author} {\bibinfo {author} {\bibfnamefont {G.}~\bibnamefont
  {S\'anchez-Sanz}}, \bibinfo {author} {\bibfnamefont {L.}~\bibnamefont
  {Seijo}}, \ and\ \bibinfo {author} {\bibfnamefont {Z.}~\bibnamefont
  {Barandiar\'an}},\ }\href@noop {} {\bibfield  {journal} {\bibinfo  {journal}
  {J. Chem. Phys.},\ }\textbf {\bibinfo {volume} {133}},\ \bibinfo {pages}
  {114509} (\bibinfo {year} {2010}{\natexlab{a}})}\BibitemShut {NoStop}%
\bibitem [{\citenamefont {S\'anchez-Sanz}\ \emph
  {et~al.}(2010){\natexlab{b}}\citenamefont {S\'anchez-Sanz}, \citenamefont
  {Seijo},\ and\ \citenamefont {Barandiar\'an}}]{SaSeBa10b}%
  \BibitemOpen
  \bibfield  {author} {\bibinfo {author} {\bibfnamefont {G.}~\bibnamefont
  {S\'anchez-Sanz}}, \bibinfo {author} {\bibfnamefont {L.}~\bibnamefont
  {Seijo}}, \ and\ \bibinfo {author} {\bibfnamefont {Z.}~\bibnamefont
  {Barandiar\'an}},\ }\href@noop {} {\bibfield  {journal} {\bibinfo  {journal}
  {J. Chem. Phys.},\ }\textbf {\bibinfo {volume} {133}},\ \bibinfo {pages}
  {114506} (\bibinfo {year} {2010}{\natexlab{b}})}\BibitemShut {NoStop}%
\bibitem [{\citenamefont {Reid}\ \emph {et~al.}(2010)\citenamefont {Reid},
  \citenamefont {Hu}, \citenamefont {Frank}, \citenamefont {Duan},
  \citenamefont {Xia},\ and\ \citenamefont {Yin}}]{ReHuFrDuXiYi10}%
  \BibitemOpen
  \bibfield  {author} {\bibinfo {author} {\bibfnamefont {M.~F.}\ \bibnamefont
  {Reid}}, \bibinfo {author} {\bibfnamefont {L.}~\bibnamefont {Hu}}, \bibinfo
  {author} {\bibfnamefont {S.}~\bibnamefont {Frank}}, \bibinfo {author}
  {\bibfnamefont {C.~K.}\ \bibnamefont {Duan}}, \bibinfo {author}
  {\bibfnamefont {S.}~\bibnamefont {Xia}}, \ and\ \bibinfo {author}
  {\bibfnamefont {M.}~\bibnamefont {Yin}},\ }\href@noop {} {\bibfield
  {journal} {\bibinfo  {journal} {Eur. J. Inorg. Chem},\ }\textbf {\bibinfo
  {volume} {2010}},\ \bibinfo {pages} {2649 } (\bibinfo {year}
  {2010})}\BibitemShut {NoStop}%
\bibitem [{\citenamefont {Grinberg}\ and\ \citenamefont
  {Mahlik}(2008)}]{GrMa08}%
  \BibitemOpen
  \bibfield  {author} {\bibinfo {author} {\bibfnamefont {M.}~\bibnamefont
  {Grinberg}}\ and\ \bibinfo {author} {\bibfnamefont {S.}~\bibnamefont
  {Mahlik}},\ }\href@noop {} {\bibfield  {journal} {\bibinfo  {journal} {J.
  Non-Cryst. Sol.},\ }\textbf {\bibinfo {volume} {354}},\ \bibinfo {pages}
  {4163} (\bibinfo {year} {2008})}\BibitemShut {NoStop}%
\bibitem [{\citenamefont {Mahlik}\ \emph {et~al.}(2009)\citenamefont {Mahlik},
  \citenamefont {Grinberg}, \citenamefont {Cavalli}, \citenamefont
  {Bettinelli},\ and\ \citenamefont {Boutinaud}}]{MaGrCaBeBo09}%
  \BibitemOpen
  \bibfield  {author} {\bibinfo {author} {\bibfnamefont {S.}~\bibnamefont
  {Mahlik}}, \bibinfo {author} {\bibfnamefont {M.}~\bibnamefont {Grinberg}},
  \bibinfo {author} {\bibfnamefont {E.}~\bibnamefont {Cavalli}}, \bibinfo
  {author} {\bibfnamefont {M.}~\bibnamefont {Bettinelli}}, \ and\ \bibinfo
  {author} {\bibfnamefont {P.}~\bibnamefont {Boutinaud}},\ }\href@noop {}
  {\bibfield  {journal} {\bibinfo  {journal} {J. Phys: Condens. Matter},\
  }\textbf {\bibinfo {volume} {21}},\ \bibinfo {pages} {105401} (\bibinfo
  {year} {2009})}\BibitemShut {NoStop}%
\bibitem [{\citenamefont {McClure}\ and\ \citenamefont
  {Pedrini}(1985)}]{McPe85}%
  \BibitemOpen
  \bibfield  {author} {\bibinfo {author} {\bibfnamefont {D.~S.}\ \bibnamefont
  {McClure}}\ and\ \bibinfo {author} {\bibfnamefont {C.}~\bibnamefont
  {Pedrini}},\ }\href@noop {} {\bibfield  {journal} {\bibinfo  {journal} {Phys.
  Rev. B},\ }\textbf {\bibinfo {volume} {32}},\ \bibinfo {pages} {8465}
  (\bibinfo {year} {1985})}\BibitemShut {NoStop}%
\bibitem [{\citenamefont {Dorenbos}(2003)}]{Do03a}%
  \BibitemOpen
  \bibfield  {author} {\bibinfo {author} {\bibfnamefont {P.}~\bibnamefont
  {Dorenbos}},\ }\href@noop {} {\bibfield  {journal} {\bibinfo  {journal} {J.
  Phys. Condensed Matter},\ }\textbf {\bibinfo {volume} {15}},\ \bibinfo
  {pages} {2645} (\bibinfo {year} {2003})}\BibitemShut {NoStop}%
\bibitem [{\citenamefont {Moine}\ \emph {et~al.}(1989)\citenamefont {Moine},
  \citenamefont {Courtois},\ and\ \citenamefont {Pedrini}}]{MoCoPe89}%
  \BibitemOpen
  \bibfield  {author} {\bibinfo {author} {\bibfnamefont {B.}~\bibnamefont
  {Moine}}, \bibinfo {author} {\bibfnamefont {B.}~\bibnamefont {Courtois}}, \
  and\ \bibinfo {author} {\bibfnamefont {C.}~\bibnamefont {Pedrini}},\
  }\href@noop {} {\bibfield  {journal} {\bibinfo  {journal} {J. de Physique},\
  }\textbf {\bibinfo {volume} {50}},\ \bibinfo {pages} {2105} (\bibinfo {year}
  {1989})}\BibitemShut {NoStop}%
\bibitem [{\citenamefont {Moine}\ \emph {et~al.}(1991)\citenamefont {Moine},
  \citenamefont {Courtois},\ and\ \citenamefont {Pedrini}}]{MoCoPe91}%
  \BibitemOpen
  \bibfield  {author} {\bibinfo {author} {\bibfnamefont {B.}~\bibnamefont
  {Moine}}, \bibinfo {author} {\bibfnamefont {B.}~\bibnamefont {Courtois}}, \
  and\ \bibinfo {author} {\bibfnamefont {C.}~\bibnamefont {Pedrini}},\
  }\href@noop {} {\bibfield  {journal} {\bibinfo  {journal} {J. Luminescence},\
  }\textbf {\bibinfo {volume} {48-49}},\ \bibinfo {pages} {501} (\bibinfo
  {year} {1991})}\BibitemShut {NoStop}%
\bibitem [{\citenamefont {Pedrini}\ \emph {et~al.}(2007)\citenamefont
  {Pedrini}, \citenamefont {Joubert},\ and\ \citenamefont
  {McClure}}]{PeJoMc07}%
  \BibitemOpen
  \bibfield  {author} {\bibinfo {author} {\bibfnamefont {C.}~\bibnamefont
  {Pedrini}}, \bibinfo {author} {\bibfnamefont {M.~F.}\ \bibnamefont
  {Joubert}}, \ and\ \bibinfo {author} {\bibfnamefont {D.~S.}\ \bibnamefont
  {McClure}},\ }\href@noop {} {\bibfield  {journal} {\bibinfo  {journal} {J.
  Luminescence},\ }\textbf {\bibinfo {volume} {125}},\ \bibinfo {pages} {230}
  (\bibinfo {year} {2007})}\BibitemShut {NoStop}%
\bibitem [{\citenamefont {Izeddin}\ \emph {et~al.}(2007)\citenamefont
  {Izeddin}, \citenamefont {Klik}, \citenamefont {Vinh}, \citenamefont
  {Bresler},\ and\ \citenamefont {Gregorkiewicz}}]{IzKlViBrGr07}%
  \BibitemOpen
  \bibfield  {author} {\bibinfo {author} {\bibfnamefont {I.}~\bibnamefont
  {Izeddin}}, \bibinfo {author} {\bibfnamefont {M.~A.~J.}\ \bibnamefont
  {Klik}}, \bibinfo {author} {\bibfnamefont {N.~Q.}\ \bibnamefont {Vinh}},
  \bibinfo {author} {\bibfnamefont {M.~S.}\ \bibnamefont {Bresler}}, \ and\
  \bibinfo {author} {\bibfnamefont {T.}~\bibnamefont {Gregorkiewicz}},\
  }\href@noop {} {\bibfield  {journal} {\bibinfo  {journal} {Phys. Rev.
  Lett.},\ }\textbf {\bibinfo {volume} {99}},\ \bibinfo {pages} {077401}
  (\bibinfo {year} {2007})}\BibitemShut {NoStop}%
\bibitem [{\citenamefont {Hayes}\ \emph {et~al.}(1973)\citenamefont {Hayes},
  \citenamefont {Wiltshire}, \citenamefont {Manthey},\ and\ \citenamefont
  {{McClure}}}]{HaWiMaMc73}%
  \BibitemOpen
  \bibfield  {author} {\bibinfo {author} {\bibfnamefont {W.}~\bibnamefont
  {Hayes}}, \bibinfo {author} {\bibfnamefont {M.~C.~K.}\ \bibnamefont
  {Wiltshire}}, \bibinfo {author} {\bibfnamefont {W.~J.}\ \bibnamefont
  {Manthey}}, \ and\ \bibinfo {author} {\bibfnamefont {D.~S.}\ \bibnamefont
  {{McClure}}},\ }\href@noop {} {\bibfield  {journal} {\bibinfo  {journal} {J.
  Phys. C},\ }\textbf {\bibinfo {volume} {6}},\ \bibinfo {pages} {L273}
  (\bibinfo {year} {1973})}\BibitemShut {NoStop}%
\bibitem [{\citenamefont {Cowan}(1981)}]{Cow81}%
  \BibitemOpen
  \bibfield  {author} {\bibinfo {author} {\bibfnamefont {R.~D.}\ \bibnamefont
  {Cowan}},\ }\href@noop {} {\emph {\bibinfo {title} {The Theory of Atomic
  Structure and Spectra}}}\ (\bibinfo  {publisher} {U. California},\ \bibinfo
  {address} {Berkeley},\ \bibinfo {year} {1981})\BibitemShut {NoStop}%
\bibitem [{\citenamefont {Le{\'s}niak}(1990)}]{Les90}%
  \BibitemOpen
  \bibfield  {author} {\bibinfo {author} {\bibfnamefont {K.}~\bibnamefont
  {Le{\'s}niak}},\ }\href@noop {} {\bibfield  {journal} {\bibinfo  {journal}
  {J. Phys.: Condens. Matter},\ }\textbf {\bibinfo {volume} {2}},\ \bibinfo
  {pages} {5563} (\bibinfo {year} {1990})}\BibitemShut {NoStop}%
\bibitem [{\citenamefont {Henderson}\ and\ \citenamefont
  {Imbusch}(1989)}]{HI89}%
  \BibitemOpen
  \bibfield  {author} {\bibinfo {author} {\bibfnamefont {B.}~\bibnamefont
  {Henderson}}\ and\ \bibinfo {author} {\bibfnamefont {G.~F.}\ \bibnamefont
  {Imbusch}},\ }\href@noop {} {\emph {\bibinfo {title} {Optical spectroscopy of
  inorganic solids}}}\ (\bibinfo  {publisher} {Clarendon Press, Oxford},\
  \bibinfo {year} {1989})\BibitemShut {NoStop}%
\end{thebibliography}
\end{document}